\documentclass[12pt]{iopart}

\usepackage{iopams}
\usepackage{graphicx}
\begin{document}

\title[Gap structure in the electron-doped Iron-Arsenide Superconductor]{Gap structure in the electron-doped Iron-Arsenide Superconductor Ba(Fe$_{0.92}$Co$_{0.08}$)$_{2}$As$_{2}$:\\ low-temperature specific heat study}

\author{K. Gofryk$^{1}$, A. S. Sefat$^{2}$, E. D. Bauer$^{1}$, M. A. McGuire$^{2}$, B. C. Sales$^{2}$, D. Mandrus$^{2}$, J. D. Thompson$^{1}$ and F. Ronning$^{1}$}

\address{$^{1}$Condensed Matter and Thermal Physics, Los Alamos National Laboratory, Los Alamos, New Mexico 87545, USA\\
$^{2}$Materials Science and Technology Division, Oak Ridge National Laboratory, Oak Ridge, Tennessee 37831, USA}
\ead{gofryk@lanl.gov}

\begin{abstract}
We report the field and temperature dependence of the low-temperature specific heat down to 400~mK and in magnetic fields up to 9~T of the electron-doped Ba(Fe$_{0.92}$Co$_{0.08}$)$_{2}$As$_{2}$ superconductor.
Using the phonon specific heat obtained from pure BaFe$_{2}$As$_{2}$ we find the normal state Sommerfeld coefficient to be 18~mJ/mol~K$^{2}$ and a condensation energy of 1.27 J/mol. The temperature dependence of the electronic specific heat clearly indicate the presence of the low-energy excitations in the system. The magnetic field variation of field-induced specific heat cannot be described by single clean $s$- or $d$-wave models. Rather, the data require an anisotropic gap scenario which may or may not have nodes. We discuss the implications of these results.
\end{abstract}

\pacs{71.20.Eh, 71.55.Ak, 72.15.Eb, 72.15.Jf, 75.50.Pp}
\vspace{2pc}
\submitto{\NJP}
\maketitle

\section{Introduction}

The recent discovery of superconductivity in Fe-based pnictides $R$FeAsO\cite{Kamihara,chen} ($R$-rare earth) has created a new era in superconductivity research and stimulated a great interest in these compounds\cite{norman}. Subsequently, other types of superconducting materials containing FeAs layers were discovered such as binary chalocgenides Fe$_{1+x}$Se\cite{hsu,11b}, so called $"111"$ compounds\cite{111a,111b} LiFeAs or NaFeAs and $122$-systems $A$Fe$_{2}$As$_{2}$ where $A$ is an alkaline earth\cite{122a,122b,122c}. BaFe$_{2}$As$_{2}$, prototypical member of the latter family, crystalizes with the tetragonal ThCr$_{2}$Si$_{2}$-structure type and at ambient pressure exhibits structural and spin-density-wave (SDW) transitions at about 140~K\cite{122Ba}. Suppression of the SDW state by either applied pressure\cite{alireza} or chemical doping\cite{122a,122c} results in superconductivity. Despite a large theoretical and experimental effort (see Ref.\cite{th,ex}) to understand the nature of the superconductivity in these materials there are still many open questions that have not been resolved such as the pairing mechanism and the symmetry of the order parameter. Moreover, the experimental results reported so far are often contradictory, ranging from nodal to fully gapped isotropic superconductivity. Even within the Co doped BaFe$_{2}$As$_{2}$ family the situation is unclear. While surface sensitive measurements such as ARPES\cite{terashima} and STM\cite{yin} claim fairly isotropic gap values, penetration depth\cite{gordon}, $\mu$SR\cite{williams}, NMR\cite{ning}, thermal conductivity\cite{tanatar, dong, machida}, specific heat\cite{mu}, and Raman scattering\cite{muschler} argue that an anisotropic gap is necessary, although the details vary between these measurements as well. Often, the superconducting gap structure is discussed in terms of the $s\pm$ model, with a sign reversal of the order parameter between different Fermi surface sheets\cite{s1,s2,s3}. However, the large sample, family, and doping dependence may favor scenarios where the low energy excitations, possibly nodal, depend strongly on the particular sample being studied and the probe used to investigate them (e.g. Ref.\cite{kuroki,wang,graser}).

In this paper we present results of our detailed studies of the specific heat of the electron-doped Ba(Fe$_{0.92}$Co$_{0.08}$)$_{2}$As$_{2}$ superconductor. By subtracting the lattice contribution (obtained from measurements on non-superconducting samples), we can extract the full electronic $T$-dependence. The temperature and magnetic field dependent data imply an anisotropic gap structure.\\

\section{Experimental details}

The large single crystals of Ba(Fe$_{0.92}$Co$_{0.08}$)$_{2}$As$_{2}$ have been grown out of FeAs flux with the typical size of about 2$\times$1.5$\times$0.2 mm$^{3}$. The samples crystalize as well-formed plates with the [001] direction perpendicular to the plane of the crystals. The doping level was determined by microprobe analysis. More details about the synthesis and characterization of the samples may be found in Ref.\cite{122b}. Based on electrical resistivity measurements, $T_{c}$ was established to be 20~K (zero resistance), in agreement with the heat capacity results presented here.
The heat capacity was measured down to 400~mK and in magnetic fields up to 9~T using a thermal relaxation method implemented in a Quantum Design PPMS-9 device. All data measured in field were field cooled.\\

\section{Results and discussion}

The temperature dependence of the specific heat of Ba(Fe$_{0.92}$Co$_{0.08}$)$_{2}$As$_{2}$ is shown in Fig.~1. As can be seen from the figure, a pronounced anomaly of specific heat is observed at $T_{c}$. A magnetic field of 9 T strongly suppresses the anomaly and moves it to lower temperatures. In general, the total specific heat of any system is the sum of several different excitations:

\begin{equation}
C_{tot}(T)=C_{el}(T)+C_{ph}(T)+C_{mag}(T)+ ...
\end{equation}

where $C_{el}(T)$, $C_{ph}(T)$ and $C_{mag}(T)$ describe electronic, lattice and magnetic contributions to the total specific heat, respectively.

\begin{figure}[ht]
\centering
\includegraphics[width=0.7\textwidth]{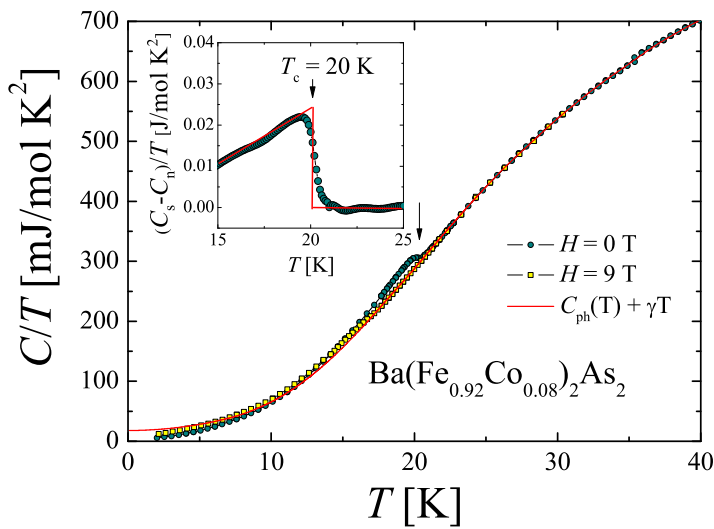}.
\caption{(Color online) The heat capacity of Ba(Fe$_{0.92}$Co$_{0.08}$)$_{2}$As$_{2}$ measured in 0 T (circles) and 9 T (squares). The solid line describes the normal state specific heat (see text). The inset shows the difference between measured $C/T$ and the phonon contribution. The $T_{c}$ obtained by entropy balance construction is 20~K.}\label{fig1}
\end{figure}

In order to estimate the phonon contribution in our system, we assume that the phonon part of the specific heat is independent of doping. Thus, we determine the phonon specific heat from the parent compound BaFe$_{2}$As$_{2}$. BaFe$_{2}$As$_{2}$ exhibits a spin-density-wave transition at about 140~K so one may expect the presence of a magnetic contribution to the low-temperatures specific heat in the system. However, recent inelastic neutron scattering experiments show that, in the ordered state, spin-wave excitations have a gap of about 10~meV ($\Delta~\approx$~116~K)\cite{matan}. Consequently, at temperatures below 40~K, $C_{mag}$ is negligible, and we separate the contributions to the specific heat of the parent compound as $C$~=~$\gamma T$ + $C_{ph}$.

Thus, to describe the experimental data of Ba(Fe$_{0.92}$Co$_{0.08}$)$_{2}$As$_{2}$ above $T_{c}$~=~20~K, we adjust gamma to obtain the best agreement between the data and $C$~=~$\gamma$T + $C_{ph}$. There are four points which give us confidence that our determination of $\gamma$ and $C_{ph}$ for the doped compound is reasonable. (i) the good agreement of $C$~=~$\gamma T$ + $C_{ph}$ above 20~K (see Fig.1). (ii) $\gamma$ which we obtain in this procedure provides accurate entropy balance for the electronic specific heat in the superconducting state below $T_{c}$. (iii) the condensation energy which we obtain from the resulting analysis is quantitatively consistent with other measurements (see below). (iv) Calculations and inelastic x-ray scattering measurements indicate that the phonons below 10~meV are independent of dopping\cite{Reznick,Reznick2}.

The value of the normal state electronic specific heat may be compared with the density of states at the Fermi level calculated for pure BaFe$_{2}$As$_{2}$. Within a single band model approximation, $\gamma$~=~18~mJ/mol~K$^{2}$ gives $N$($E_{F}$) to be 7.64~eV$^{-1}$/f.u. Using LDA approximation together with GGA-PBE\cite{ma} or general potential in LAPW method\cite{singh} the calculated values of $N$($E_{F}$), for the parent BaFe$_{2}$As$_{2}$, are 3.93 and 3.06~eV$^{-1}$/f.u., respectively. Thus the mass renormalization is roughly a factor of 2, consistent with mass renormalization determined by optics\cite{MMQ} and ARPES\cite{lu}.

At 20~K the specific heat exhibits a jump $\Delta C/T_{c}$~=~24~mJ/mol~K$^{2}$ (see the inset in Fig.1) being consistent with previous reports (Ref.\cite{122b,mu,budko,chu}). It is also similar to $\Delta C/T_{c}$~=~28~mJ/mol~K$^{2}$ obtained for Ba$_{0.6}$K$_{0.4}$Fe$_{2}$As$_{2}$\cite{mu,kant}. The size of the jump at $T_{c}$ depends on the details of the superconducting state. Taking $\gamma$~=~18~mJ/mol~K$^{2}$ the ratio $\Delta C/T_{c}\gamma$~=~1.33 is very close to, albeit smaller than, the weak-coupling BCS value 1.43.

\begin{figure}[htbp]
\centering
\includegraphics[width=0.7\textwidth]{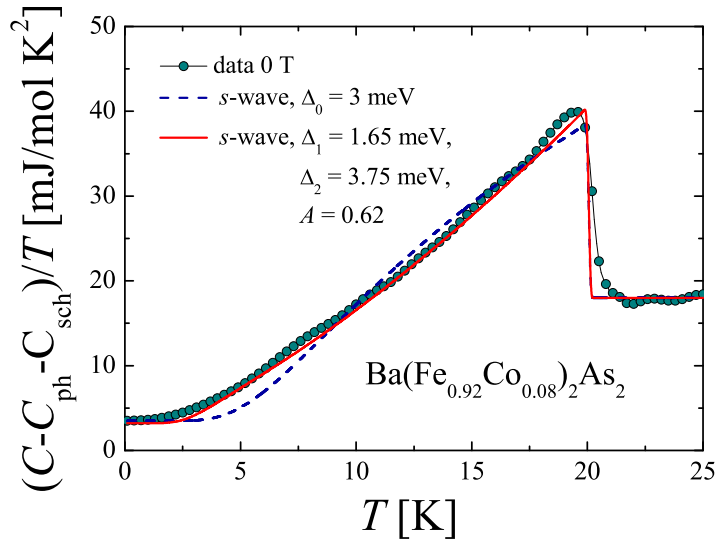}
\caption{(Color online) Temperature dependence of the electronic specific of Ba(Fe$_{0.92}$Co$_{0.08}$)$_{2}$As$_{2}$. The dashed and solid lines are theoretical curves based on BCS theory (Eq.3) with a one and two s-wave gaps, respectively (see text).}\label{fig2}

\end{figure}

Using the normal state specific heat, we can extract the condensation energy and relate it to the thermodynamic
critical field. This quantity can be obtained by integrating the entropy difference:

\begin{equation}
U=\int \left[S_{n}(T)-S_{s}(T)\right]dT
\end{equation}

where $S_{n}$ and $S_{s}$ denote entropy at the normal and superconducting state respectively. In the case of Ba(Fe$_{0.92}$Co$_{0.08}$)$_{2}$As$_{2}$ this analysis give $U$~=~1.27~J/mol and the thermodynamic
critical field $H_{c}$~=~0.23~T. Using a penetration depth $\lambda$~=~325~nm obtained from MFM\cite{R1} at a slightly different doping level and consistent with $\mu$SR results\cite{R2} and coherence length $\xi$~=~27.6 obtained by STS\cite{yin} gives a Ginzburg-Landau parameter $K~=~\frac{\lambda}{\xi}$~=~118. Using this value and our thermodynamic critical field $H_{c}$ we obtain $H_{c2}$~=~38~T, from the expression $H_{c2}$~=~$\sqrt{2}KH_{c}$, in reasonable agreement with the published value of about 40~T\cite{ni} and the value of 39~T obtained from the slope of the upper critical field measured by specific heat and the expression $H_{c2}$=0.69$\frac{dH_{c2}}{dT_{c}}T_{c}$\cite{WHH}. This provides additional confidence in our phonon substraction

We begin our investigation on the possible symmetry of the superconducting gap by examining the temperature dependence of the electronic specific heat. Fig.2 displays the non-lattice part of the specific heat of Ba(Fe$_{0.92}$Co$_{0.08}$)$_{2}$As$_{2}$ obtained by subtracting the phonon contribution together with a small Schottky contribution below 1~K (see below). At low temperatures, a sizeable residual specific heat coefficient $\gamma_{0}$~=~3.7~mJ/mol~K$^{2}$ is observed in this system. Similar behavior has also been reported in Ba$_{0.6}$K$_{0.4}$Fe$_{2}$As$_{2}$ ($\gamma_{0}$~=~7.7~mJ/mol~K$^{2}$)\cite{mu2} and Ba(Fe$_{1-x}$Co$_{x}$)$_{2}$As$_{2}$ ($\gamma_{0}$~$\approx$~3~mJ/mol~K$^{2}$ for optimal doped samples)\cite{mu}. Interestingly, the sizeable value of the residual specific heat coefficient also has been observed in superconducting cuprates\cite{cuprates,hussey}. For Ba(Fe$_{0.92}$Co$_{0.08}$)$_{2}$As$_{2}$, $\gamma_{0}$~=~3.7~mJ/mol~K$^{2}$ amounts to 20~\% of $\gamma$. We rule out that $\gamma_{0}$ originates from a non-superconducting portion of the sample based on the fact that x-ray analysis limits impurity phases to less than 5\%, and we observe full diamagnetic shielding from magnetization measurements. Alternative explanations for the origin of the residual $\gamma_{0}$ include pair breaking effects of an unconventional superconductor\cite{kh}, crystallographic defects or spin glass behavior.

\begin{figure}[htbp]
\centering
\includegraphics[width=0.7\textwidth]{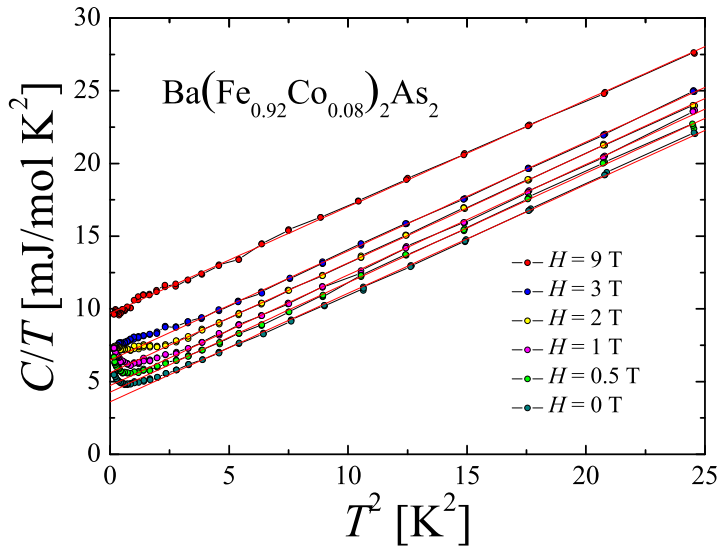}
\caption{(Color online) The heat capacity of Ba(Fe$_{0.92}$Co$_{0.08}$)$_{2}$As$_{2}$ plotted as $C/T$ vs $T^{2}$ with magnetic field applied along the c-axis.}\label{fig3}

\end{figure}

To fit the $C_{el}/T$ data in figure 2 we use the BCS expression for specific heat:

\begin{equation}
C_{BCS}=t\frac{d}{dt}\int_{0}^{\infty}dy\left(-\frac{6\gamma\Delta_{0}}{k_{B}\pi}\right)\left[flnf+\left(1-f\right)ln\left(1-f\right)\right]
\end{equation}

where $t~=~\frac{T}{T_{c}}$, $f$ is the Fermi function $f~=~\frac{1}{e^{\frac{E}{k_{B}T}}+1}$, $E~=~\sqrt{\epsilon^{2}+\Delta^{2}}$ and $y~=~\frac{\epsilon}{\Delta}$ (see Ref.\cite{tinkham}).

The results of our analysis using a single $s$-wave gap\cite{suhl} (dashed blue line) and two separate $s$-wave gaps (solid red line) are shown in Fig.2. In an $s$-wave model a residual linear term must be an extrinsic contribution to the specific heat, and hence we subtracted $\gamma_{0}$ from $\gamma$ for the purposes of these fits. The normal state Sommerfeld coefficient $\gamma$~=~(18~-~3.7)~mJ/mol~K$^{2}$ and $T_{c}$~=~20~K were held fixed during the fits. The gap value obtained from the single gap fit is 3~meV and may be compared with $\Delta_{0}$~=~6~meV derived for hole-doped Ba$_{0.6}$K$_{0.4}$Fe$_{2}$As$_{2}$ with $T_{c}~\approx$~37~K\cite{mu2}. Taking $\Delta_{0}$~=~3~meV and $T_{c}$~=~20~K gives $\Delta_{0}/(k_{B}T_{c})$~=~1.74, close to the weak coupling value of 1.76.
However, as can be seen from Fig.2 the single gap fit does not describe the data sufficiently and clearly indicates the presence of low energy excitations in the system below 8~K. A much better description is obtained by fitting the data to $C$~=~(1-A)$C_{BCS}$($\Delta_{1}$)~+~$AC_{BCS}$($\Delta_{2}$) which gives the solid red line and the parameters $\Delta_{1}$~=~1.65~meV, $\Delta_{2}$~=~3.75~meV and $A$~=~0.62. While this fit provides a reasonable description of the data, we emphasize that there are multiple anistropic gap descriptions that could provide a similarly good fit. From this data alone we cannot determine whether or not nodes exist.

\begin{figure}[htbp]
\centering
\includegraphics[width=0.7\textwidth]{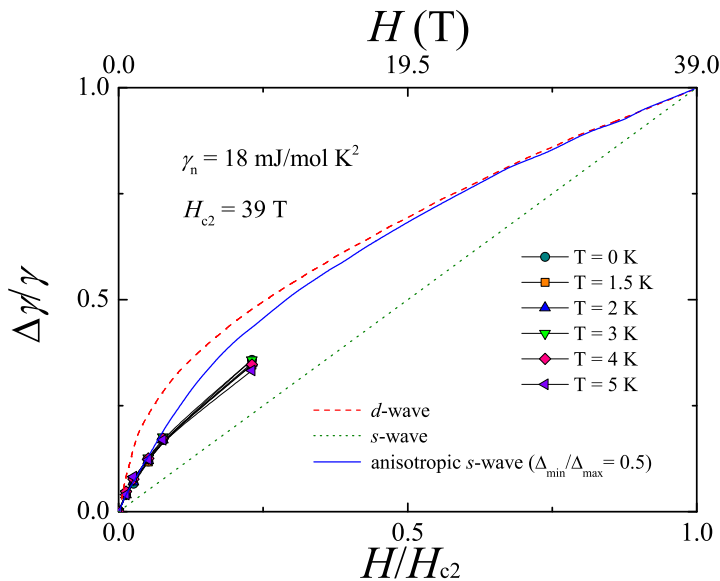}
\caption{(Color online) Field-induced change in low temperature specific heat of Ba(Fe$_{0.92}$Co$_{0.08}$)$_{2}$As$_{2}$, obtained at 0 K by extrapolating the experimental data of Fig.3 to zero temperatures (see text). The green dotted line and red dashed line represent field dependencies expected for $s$-wave and $d$-wave descriptions, respectively. The blue solid line is a theoretical curve for anisotropic $s$-wave superconductors (see text).}\label{fig4}

\end{figure}

The low-temperature specific heat of Ba(Fe$_{0.92}$Co$_{0.08}$)$_{2}$As$_{2}$ measured in several magnetic fields is presented in Fig.3. Below 1~K an upturn in $C/T$ is observed. Such behavior has been already observed in Ba(Fe$_{0.92}$Co$_{0.08}$)$_{2}$As$_{2}$ samples grown by In flux as well as in Ba$_{0.6}$K$_{0.4}$Fe$_{2}$As$_{2}$ and FeSe single crystals\cite{kim,mu2,McQueen}. The origin of the anomaly is most probably related to the presence of a small amount of magnetic impurities. With increasing magnetic field the upturn shifts to higher temperatures and transforms into a maximum, which gradually broadens and diminishes in magnitude with further rising field. Such behavior indeed resembles that expected for a degenerate ground state split due to the Zeeman effect in internal and external magnetic fields. Thus, to avoid the effects of this magnetic contribution, we rely on a linear extrapolation of the data from above 1.5~K. The red solid lines in Fig.3 display the linear tendency of low-temperature specific heat displayed as $C/T(T^{2})$. We have used it to determine the electronic specific at 0 K by extrapolating the low-temperature data to zero temperature.

The so obtained $\Delta\gamma~=~\frac{\left[C(H)-C(0)\right]}{T}$ derived at 0~K, as described above, as well as at several other temperatures is presented in Fig.4. Similar data were obtained in Ref.\cite{mu}. The data have been presented in the form $\Delta\gamma /\gamma$ vs $H/H_{c2}$ with $\gamma$~=~18~mJ/mol~K$^{2}$ and $H_{c2}$~=~39~T. In fully gapped superconductors the localized quasiparticle states in vortex cores result in $\Delta\gamma$ proportional to $H/H_{c2}$ since the number of vortices is proportional to the magnetic field. As can be seen from the figure, the specific heat rises much faster with field than expected for a simple $s$-wave gap. This could indicate an anisotropic gap or point to a field dependent coherence length\cite{sonier,ichioka}. The latter scenario, however, is unlikely to be a sole explanation as the temperature dependence indicates low-energy excitations in the system (see Fig.2). Thus, we further explore the expectation of an anisotropic gap for an explanation of the $\Delta\gamma (H)$ dependence.

It was shown theoretically\cite{nakai} by using microscopic quasiclassical theory that an anisotropic gap structure can display a significant field dependence in $\Delta\gamma$(H) . Taking into account this model the experimental data may be well described at low magnetic field (see the blue solid line in Fig.4) using the gap anisotropy ratio $\alpha$~=~0.5 ($\Delta_{min}/\Delta_{max}$~=~0.5), which is in resonable agreement with the two gap fit of the temperature dependence in Fig.2.

\begin{figure}[htbp]
\centering
\includegraphics[width=0.7\textwidth]{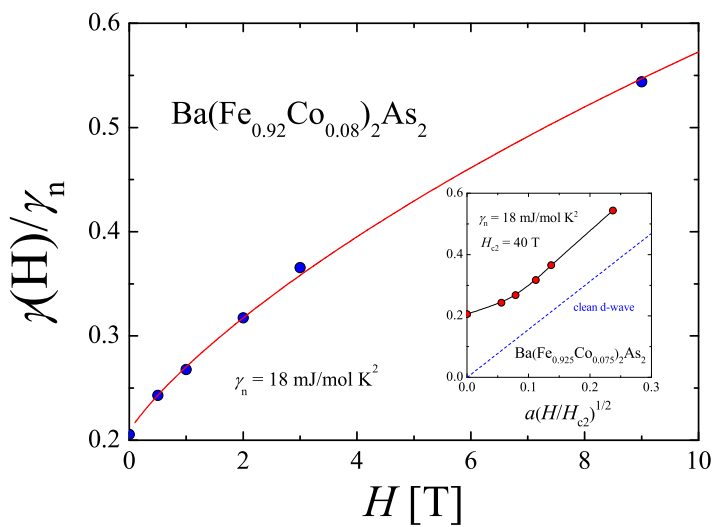}
\caption{(Color online) Normalized field-induced change in low temperature specific heat of Ba(Fe$_{0.92}$Co$_{0.08}$)$_{2}$As$_{2}$, obtained at 0 K, versus magnetic field. The red solid line is a fit of Eq.4 to the experimental data. Inset: the same data plotted versus $a\left(\frac{H}{H_{c2}}\right)^{\frac{1}{2}}$ with $a$~=~0.5 and $H_{c2}$~=~40~T.}\label{fig5}

\end{figure}

An extreme anisotropic gap is that of a nodal superconductor as occurs in the cuprates. For clean $d$-wave superconductors, it has been shown by Volovik\cite{volovik} that the quasiparticle excitation spectrum is shifted by the superfluid velocity, resulting in $\Delta\gamma~\propto~\sqrt{H/H_{c2}}$. The quasiparticles that contribute to the density of states are coming from regions far from the vortex cores and close to the nodes. This situation is presented in Fig.4 by the red dashed line. As can be seen from the figure  $\Delta\gamma$ is not increasing so strong as expected for clean $d$-wave superconductors.

Finally, we consider an additional alternative. The finite density of states $\gamma_{0}$ observed in Ba(Fe$_{0.92}$Co$_{0.08}$)$_{2}$As$_{2}$ may also be related to impurity scattering effect as expected for a dirty $d$-wave superconductor. It has been shown by K\"{u}bert and Hirschfeld\cite{kh} that in the dirty d-wave limit, $\Delta\gamma$ behaves like $H$log$H$ at the lowest fields ($H_{c1}~\leq~H~\ll~H_{c2}$). In this approach the field dependence of field-induced specific heat may be expressed by\cite{liu}:

\begin{equation}
\frac{\Delta\gamma}{\gamma_{n}}=A\left(\frac{H}{B}\right)log\left(\frac{B}{H}\right)
\end{equation}

where $A$~=~0.322$\left(\frac{\Delta_{0}}{\Gamma}\right)^{\frac{1}{2}}$, $B$~=~$\frac{\pi H_{c2}}{2a^{2}}$, $a~\approx$~0.5 and $\Delta_{0}$ and $\Gamma$ denote a maximum of the gap and impurity scattering rate, respectively. The solid line in Fig.5 is a fit of Eq.4 to the experimental data, resulting in the parameter $\frac{\Delta_{0}}{\Gamma}$~=~78. During this analysis the value of $H_{c2}$ has been fixed to 40~T. The inset in Fig.5 shows the data for Ba(Fe$_{0.92}$Co$_{0.08}$)$_{2}$As$_{2}$ together with a curve calculated for a clean $d$-wave superconductor. As may be seen the field dependence of $\gamma(H)/\gamma_{n}$ as well as the magnitude of the residual specific heat may be consistent with the dirty $d$-wave scenario. The point of these two fits (Fig.4 and 5) is to demonstrate that we can not distinguish between these two different interpretations. The quality of agreement between the two gap analysis in Fig.4 and the dirty $d$-wave analysis of Fig.5 is comparable.\\

From the above analysis, we can determine that a single isotropic s-wave gap is incapable of describing the specific heat data. The form of the anisotropic gap, however, cannot be determined by our data alone. We see that two extreme cases of multiband $s$-wave, and a dirty $d$-wave scenario are each in reasonable agreement with the data. Results from NMR\cite{ning}, thermal conductivity\cite{tanatar, dong, machida}, ARPES\cite{terashima}, penetration depth\cite{gordon}, $\mu$SR\cite{williams} and Raman\cite{muschler} also conclude that an anisotropic gap is necessary to describe their data on Co-doped BaFe$_{2}$As$_{2}$, although the extent to which varies from gapless to mild multiband behavior. A nodal gap imposed by symmetry, as in the case of the cuprates, is ruled out by the vanishingly small residual linear term of the thermal conductivity. Hence, the dirty $d$-wave analysis applied above should not be directly applicable. However, accidental nodes as anticipated in some spin-fluctuation models of the pnictides (see Ref.\cite{kuroki,wang,graser}), cannot be ruled out by the thermal conductivity results\cite{mishra} and are also consistent with our specific heat data. The accidental node scenario has the favorable aspect that the nodes could be lifted by disorder and/or doping which would help reconcile some of the seemingly contradictory results\cite{mishra2}. Further measurements as a function of doping and disorder are necessary to help elucidate the gap structure, not to mention the variations between different families and dopant atoms.\\

\section{Summary}

In summary, we have used low-temperature specific heat and its magnetic field response to explore details of the symmetry of the superconducting gap in electron-doped Ba(Fe$_{0.92}$Co$_{0.08}$)$_{2}$As$_{2}$ superconductor with $T_{c}$~=~20~K. Using the phonon part of the specific heat of pure BaFe$_{2}$As$_{2}$, we determine the normal state Sommerfeld coefficient in Ba(Fe$_{0.92}$Co$_{0.08}$)$_{2}$As$_{2}$ to be 18~mJ/mol~K$^{2}$. The temperature variation of the electronic specific heat below $T_{c}$ may be well described by the presence of two superconducting gaps, pointing to complex gap structure in the system. The field-induced low-temperature specific heat can not be explained by simple clean $s$- or $d$-wave descriptions. Its behavior also indicates a strongly anisotropic gapped superconductor.\\

\textit{Note}: During completion of this manuscript we became aware of ref.\cite{hardy} which used a similar procedure to determine the phonon contribution to the specific heat of a similar crystal. The resulting temperature dependence was analyzed within a two band model with results in good agreement with ours.

\subsection{Acknowledgments}

Work at Los Alamos National Laboratory was
performed under the auspices of the U.S. Department of
Energy, Office of Science and supported in part by the Los Alamos LDRD program. Research at Oak Ridge National Laboratory is sponsored by the Division of Material Sciences and Engineering Office of Basic Energy Sciences.

\end{document}